\documentclass[conference]{IEEEtran}

\usepackage{graphics} 
\usepackage[pdftex]{graphicx}
\usepackage[english]{babel}
\usepackage{soul}
\usepackage{epstopdf}
\DeclareGraphicsExtensions{.eps,.pdf,.jpeg,.png}
\usepackage{epsfig} 
\usepackage[compress]{cite}
\graphicspath{{./images/}}
\usepackage{array}
\usepackage{gensymb}
\usepackage{tabularx}
\usepackage{tabulary}
\newcolumntype{C}[1]{>{\centering\arraybackslash}p{#1}}
\usepackage{booktabs}
\usepackage{adjustbox}
\usepackage{float}
\usepackage[caption=false,font=normalsize,labelfont=sf,textfont=sf]{subfig}
\usepackage{filecontents}
\usepackage{amsmath}
\usepackage{color,soul}
\usepackage{wrapfig}

\begin{document}
%
\title{Augmented Reality Prosthesis Training Setup for Motor Skill Enhancement}

\author{
	\IEEEauthorblockN{Avinash~Sharma\IEEEauthorrefmark{1}, Wally~Niu\IEEEauthorrefmark{1}, Christopher~L.~Hunt\IEEEauthorrefmark{1},
				}
	\IEEEauthorblockN{Gy\"{o}rgy L\'evay\IEEEauthorrefmark{4}, Rahul~R.~Kaliki\IEEEauthorrefmark{4},
        and~Nitish~Thakor\IEEEauthorrefmark{1}\IEEEauthorrefmark{5}}

\IEEEauthorblockA{\IEEEauthorrefmark{1}Department of Biomedical Engineering, Johns Hopkins University, Baltimore, MD 21218 USA}%

\IEEEauthorblockA{\IEEEauthorrefmark{4}Infinite Biomedical Technologies, LLC, Baltimore, MD 21218 USA}%

\IEEEauthorblockA{\IEEEauthorrefmark{5}Singapore Institute for Neurotechnology, National University of Singapore, 119077 Singapore\\Email: avinash@jhmi.edu}%
}

\maketitle

\begin{abstract}
Adjusting to amputation can often time be difficult for the body. Post-surgery, amputees have to wait for up to several months before receiving a properly fitted prosthesis. In recent years, there has been a trend toward quantitative outcome measures. In this paper, we developed the augmented reality (AR) version of one such measure, the Prosthetic Hand Assessment Measure (PHAM). The AR version of the PHAM --- HoloPHAM, offers amputees the advantage to train with pattern recognition, at their own time and convenience, pre- and post-prosthesis fitting. We provide a rigorous analysis of our system, focusing on its ability to simulate reach, grasp, and touch in AR. Similarity of motion joint dynamics for reach in physical and AR space were compared, with experiments conducted to illustrate how depth in AR is perceived. To show the effectiveness and validity of our system for prosthesis training, we conducted a 10-day study with able-bodied subjects (\textit{N = 3}) to see the effect that training on the HoloPHAM had on other established functional outcome measures. A washout phase of 5 days was incorporated to observe the effect without training. Comparisons were made with standardized outcome metrics, along with the progression of kinematic variability over time. Statistically significant (\textit{p}$<$0.05) improvements were observed between pre- and post-training stages. Our results show that AR can be an effective tool for prosthesis training with pattern recognition systems, fostering motor learning for reaching movement tasks, and paving the possibility of replacing physical training. \end{abstract}

\IEEEpeerreviewmaketitle
\section{Introduction}
Due to its use in a majority of activities of daily living (ADL), the upper limb is one of the most essential body parts~\cite{kumahara2004daily}. Amputation results in the loss of both motor and sensory perceptions and with over 60\% of the amputees aging between 21 to 64 years~\cite{esquenazi2004amputation}, it is pivotal that the loss of limb should not become an impediment in ones ability to perform simple daily tasks. From simple prosthesis control training to repetitive drills for skill improvement, functional assessment during occupational therapy (OT) is an essential step prior to the fitting of an actual myoelectric prosthesis~\cite{johnson2014prosthetic}. The goal of OT is to assist amputees, through a series of training regimes, to strengthen muscle sites, improve endurance, and generally build confidence with a prosthesis.
Post-surgery training has been known to maintain the overall health of the muscles in the residual limb~\cite{malone1984immediate}, and also affect the long term success of prosthesis use~\cite{biddiss2007upper}. At the cortical level, there is evidence to suggest continued unfavorable sensory reorganization, immediately after amputation~\cite{wheaton2017neurorehabilitation}. Therefore it is desirable that the prosthetic training stage starts as early as possible. 
\begin{figure}
    \centering
        \includegraphics[width=0.5\textwidth]{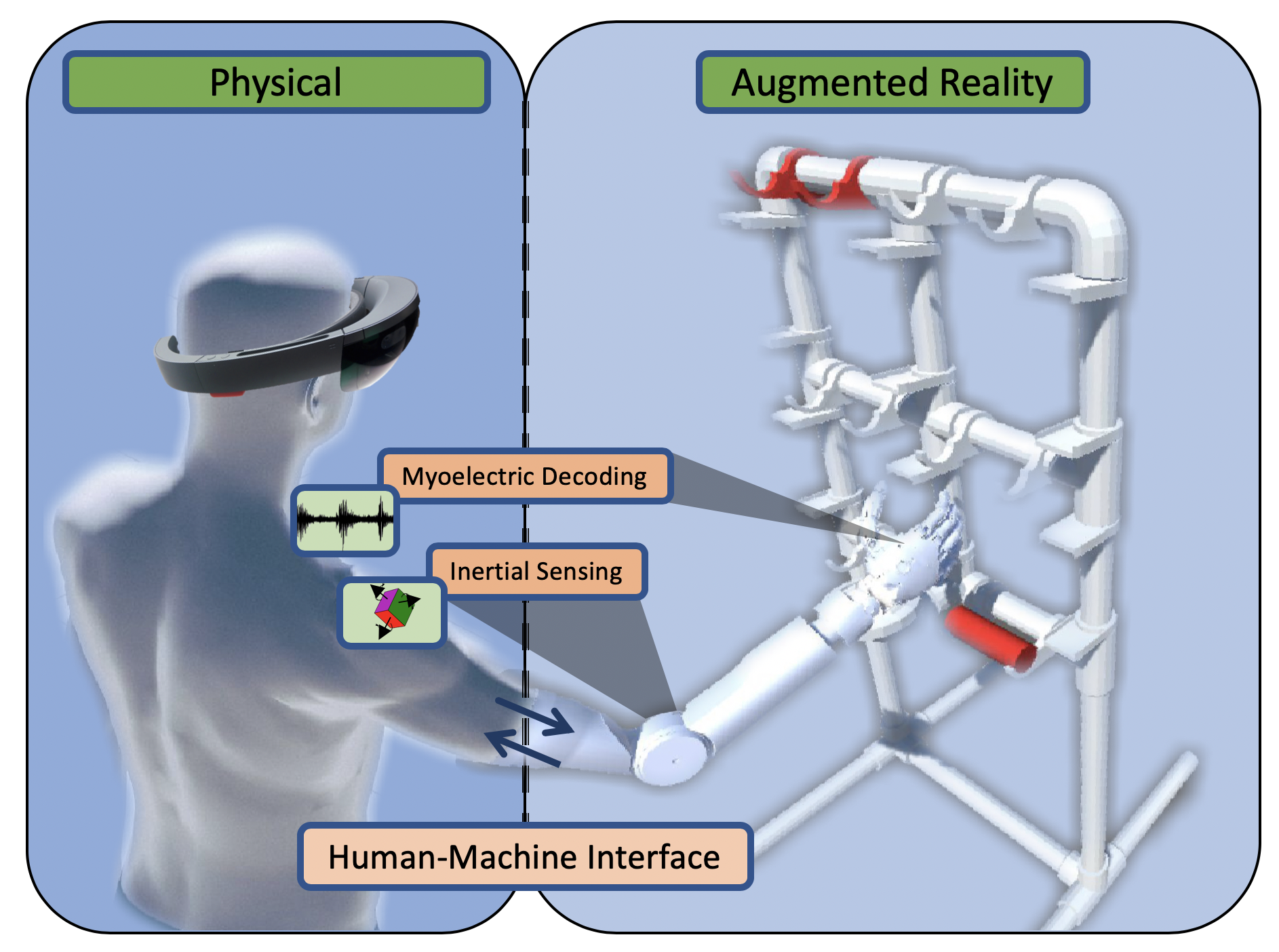}
    \caption{Our human-machine interaction system for upper limb myoelectric prosthesis control and training. Interface with the virtual modular prosthetic limb (vMPL) is through inertial sensing and myoelectric decoding that controls the kinematic dynamics and grip patterns of the limb respectively. The AR environment is projected through the Microsoft HoloLens\texttrademark~where the subject receives both visual and tactile feedback during interaction with the virtual objects.  }
    \label{fig:PHAM}
\end{figure}
While currently used functional outcome measures suffer from several drawbacks ranging from unrealistic training tasks to subjective scores~\cite{wright2009prosthetic} and limited target amputee population~\cite{prince2008comparison}, one of the major problem is the lack of consideration of practical issues that affect prosthesis use (e.g. loading and limb position effect in myoelectric decoding~\cite{hill2009functional}). 
Previous research has shown that while subjective assessment and speed of tasks are essential~\cite{bouwsema2012determining}, other measures, like limb kinematics and gaze, can provide deeper insights into the understanding of skill level~\cite{bongers2012bernstein,major2014comparison,thies2017skill,metzger2012characterization}. Major \textit{et al.}~\cite{major2014comparison} showed that kinematic variability was negatively associated with prosthesis experience and emphasized the importance of complementing motion information with existing functional clinical outcomes. Kinematic variability, specifically referring to joint kinematic trajectories, has been shown to be higher amongst prosthesis users, in comparison to able-bodied subjects~\cite{thies2017skill}. This difference, is likely due to the compensatory movements that amputees develop over time~\cite{metzger2012characterization}. That being said, there is little to no research investigating the effect that prosthesis training has on kinematic variability. To overcome some of the issues mentioned, our group designed the PHAM, an objective assessment setup allowing amputees to train over a 3-dimensional space~\cite{hunt2011pham}. However, the PHAM, like most physical outcome measures, requires a large spatial and economic investment, and is restricted to clinical settings~\cite{dromerick2008effect}.

Prosthesis training often requires repetition of certain well defined tasks~\cite{simon2012patient}. Such repetitions can be made engaging and motivating for the subjects through immersive technologies.  Virtual reality (VR) systems create an artificial environment which has a broad range of applications ranging from gaming to medicine. In upper limb prosthesis training, a range of VR systems have been developed, ranging from simplified adaptation of existing physical outcomes\cite{hargrove2007real} to myoelectric control training through posture matching\cite{simon2011target} and even virtual avatar control in custom-designed games\cite{lambrecht2011virtual,van2016task}. In a meta-analysis study, Howard \textit{et al.} showed that VR rehabilitation programs are more effective than traditional rehabilitation programs for physical outcome development~\cite{howard2017meta}. The authors suggested that user excitement and physical and cognitive fidelity could account for the increased effectiveness of virtual training programs.
Our group previously showed that myoelectric training in a virtual environment improves both consistency and distinguishability of phantom limb pattern recognition movement classes~\cite{powell2014user}. Furthermore, Hargrove \textit{et al.} recently showed that improvements gained in a physical task completions correlate linearly with improvements found in VR rehabilitation tasks\cite{hargrove2018control}. Dijk \textit{et al.} showed that virtual training can show improvements in actual prosthesis using proportional myoelectric control~\cite{van2016task}. 

While VR systems have been shown to be useful, they are not without their disadvantages. For example, VR systems require the user to sit in front of a computer or put on a head-mounted display (occluding the natural physical space), while also posing several issues like a restricted field-of-view~\cite{brooks1999s}, motion sickness~\cite{ebenholtz1992motion} and other health issues~\cite{wann1997health}. An emerging technology, that overcomes some of these shortcomings, while also creating an engaging interaction is augmented reality (AR). AR systems blend virtual objects in the physical space, while enabling interaction with these virtual objects, creating a much richer experience in comparison to VR. Several studies have used AR in stroke rehabilitation~\cite{burke2010augmented,luo2006integration,alamri2010ar}, however, use in upper limb prosthesis rehabilitation has primarily been restricted to treating phantom limb pain~\cite{dunn2017virtual,ortiz2014treatment}. Our previous preliminary work has shown that incorporating sensory feedback in AR environment for upper limb prosthesis training resulted in improvements in myoelectric control and muscular effort for able-bodied subjects~\cite{sharmaAR}. 

The objective of this study was to observe the effect of HoloPHAM (AR version of PHAM) training on standardized physical outcome measures. For the physical outcomes, the clothespin relocation test (CRT) and PHAM were used. Training was conducted using the proposed AR system (HoloPHAM) over a period of 10 days and standard metrics for comparison such as completion time and kinematic variability were studied. 
\begin{figure}[t!]
      \centering
      \framebox{\parbox{3.15in}{      \includegraphics[width=0.44\textwidth]{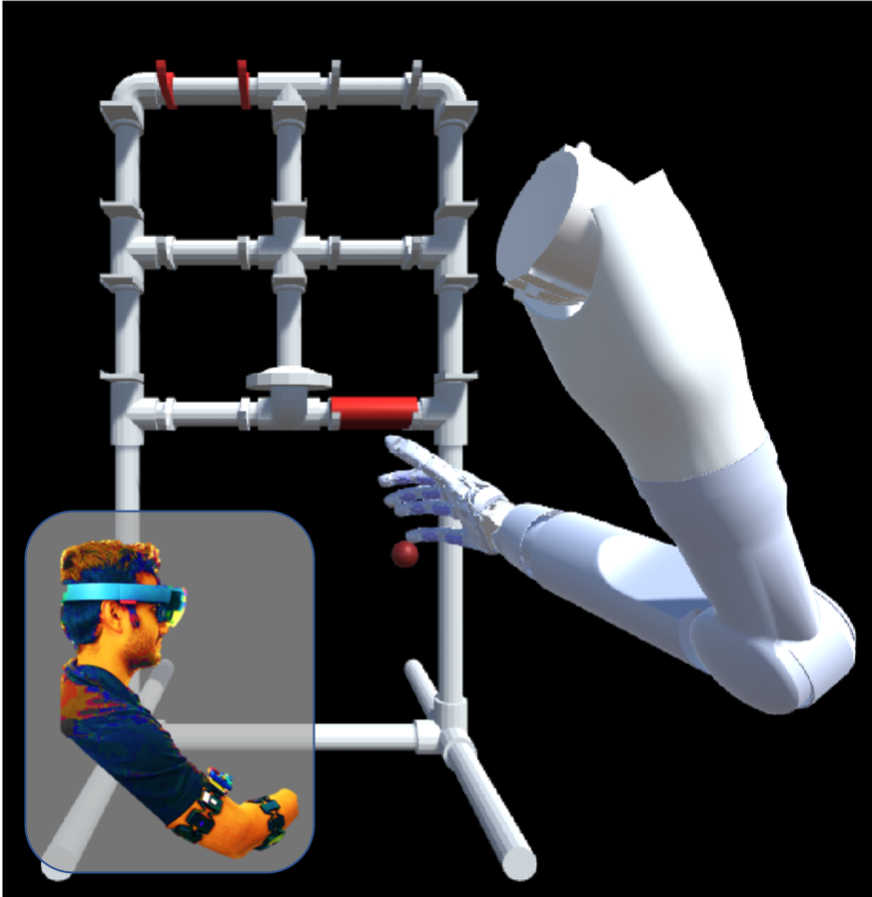}\newline \includegraphics[width=0.44\textwidth]{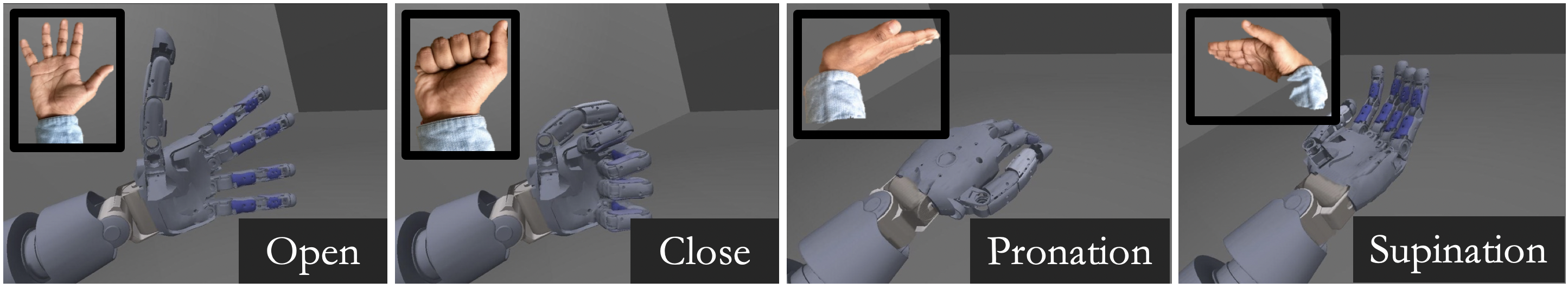}}}
      \caption{[Top] Augmented Reality scene with the vMPL and the virtual model of the PHAM (for clarity we show the scene on a dark background). (Inset) The vMPL appeared to be attached to the shoulder of the subject and move with the them in realtime using the built-in inertial sensors on the Microsoft HoloLens\texttrademark. Additional inertial sensors placed on the upper limb of the subject were used to control the motion of the vMPL which involved the vMPL reaching for a target object (red cylinder), grasp to pick and move target object, and finally drop the target object on the goal position (red tray). During pick and drop, the hand was required to be oriented to the surface normal of the objects. [Bottom] Four grip patterns (rest included, but not shown here) were integrated in our pattern recognition system. As the goal of the study was to show the effect of AR training, only a minimum number of classes were chosen, just enough to allow basic object manipulation. [Inlet] Wave out (WO) and wave in (WI) corresponded to wrist pronation (WP) and wrist supination (WS) respectively on the vMPL. In our tests, we observed that LDA could more effectively classify WO and WI, while WP and WS were often misclassified with other classes. }
      \label{fig: HoloPHAM}
   \end{figure}

\section{Methods \& Experiment}
Three able-bodied volunteers served as subjects and the experiments were conducted with their informed consent. The experiment protocol was approved by the Johns Hopkins Medicine Institutional Review Board. 

\subsection{Functional Outcome Measures}
To account for reach and grasp motions that are involved in several ADL, measures that evaluate the aspects of hand, wrist and elbow functions were used. The CRT has been an established outcome measure that accounts for precise myoelectric control as well as the coordinated movement of the upper limb joints~\cite{kyberd2018characterisation}. The PHAM has been shown to incorporate a range of motions over a 3-dimensional space, accounting for several DOFs and grasping patterns~\cite{hunt2011pham}. Since the goal of our study was to show training in AR can show improvement in performance for such movements, an AR version of the PHAM setup was developed. 

The AR version of the PHAM --- HoloPHAM, allows subjects to perform the same set of tasks as they could with the physical version. The PHAM consists of a windowpane structure, with horizontal and vertical pipes where objects of different types are required to be moved around a large activity envelope~\cite{hunt2011pham}.
While involving subtle, orientation-dependent control, the PHAM also requires reach and grasp motion. With a greater range of activity, while also preserving the fundamental aspects of existing physical prosthesis assessments, the PHAM provides a technique for evaluating both quantitative and qualitative prosthesis functionality~\cite{hunt2011pham}.
\begin{figure}[t!]
      \centering
          \includegraphics[width=0.44\textwidth]{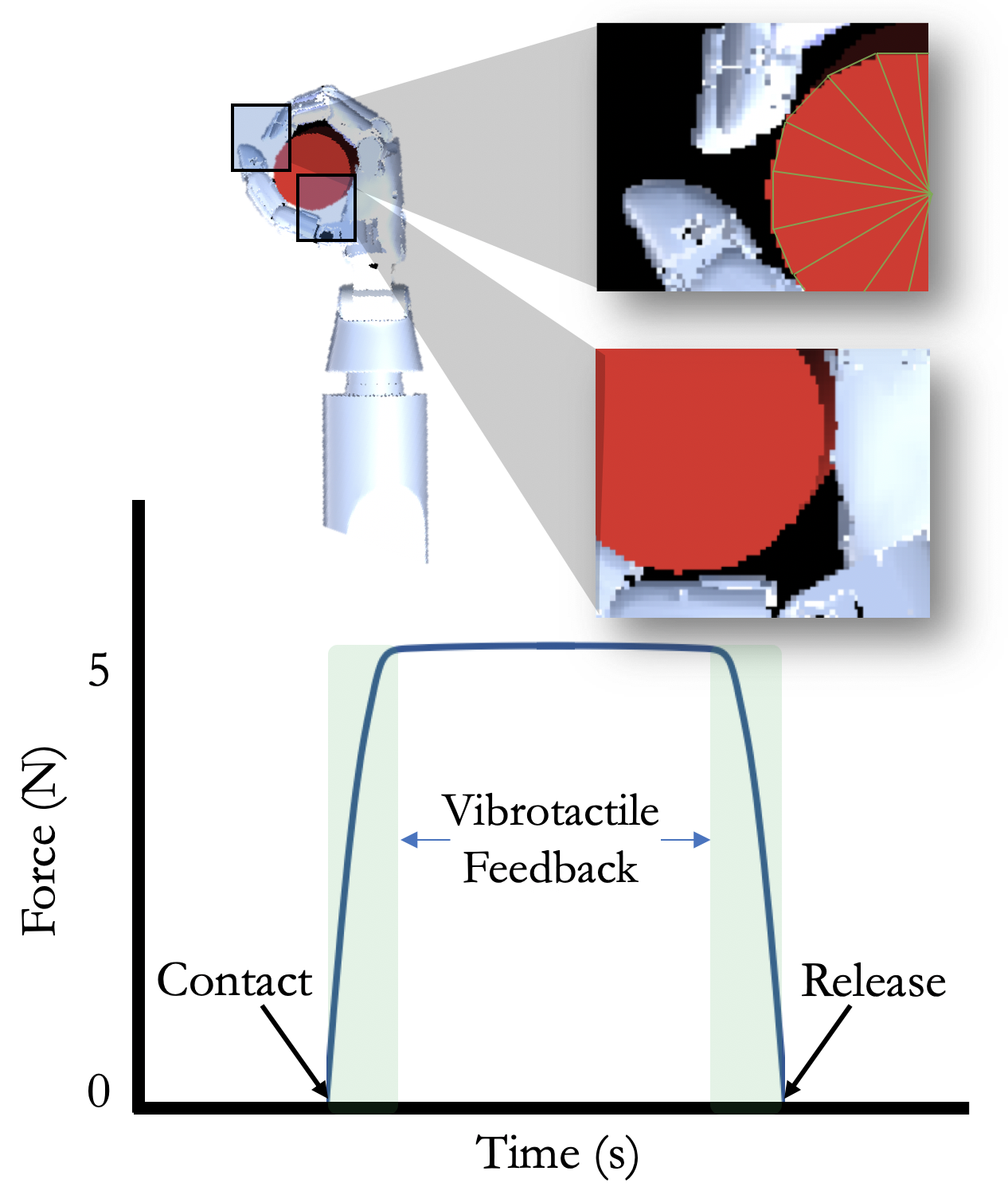} 
       \caption{[Top] Interaction between virtual objects were defined as rigid body collisions with a dense mesh collider (denoted as green lines inside the cylinder) to prevent penetration of colliding objects. We show a zoomed in image of the contact to illustrate the same. [Bottom] To validate this, force values ($F$) were calculated from the impulse ($J$) during contact and release. The impulse values ($J$) were recorded from the fingertips of the vMPL and were differentiated with time to get force values ($J = \int F dt$, here $dt$ represents the frame rate in Unity3D). Touch information was provided to the subject through a vibrotactile feedback with a gentle vibration provided by the Myo~\texttrademark~armband. }
      \label{fig:touch}
   \end{figure} 
The HoloPHAM setup comprised of a virtual model of the PHAM, viewed through the Microsoft HoloLens\texttrademark. The virtual model could be placed at any convenient location within the room for subjects to train with. Subjects interacted with the virtual objects on the HoloPHAM using a virtual version of the Modular Prosthetic Limb (vMPL, Applied Physics Laboratory, Laurel, MD). The vMPL was also projected through the HoloLens\texttrademark. All AR assets were developed in Unity3D (Unity Technologies, San Francisco) (see Fig. \ref{fig: HoloPHAM}) using the Mixed Reality Toolkit (Microsoft, Redmond, WA). The HoloLens\texttrademark asset was imported into the scene and the vMPL asset was attached to the main camera, in a configuration, so as to allow subjects to freely move around the room while also allowing a first person view of the vMPL. All objects in the scene were defined as rigid bodies with the mass of the geometric primitive objects set to $1/10$ of the mass of the vMPL. Mass of the HoloPHAM frame and the vMPL were set to be equal. These properties were set to enable rigid body collision between the interacting objects (see Fig. \ref{fig:touch}). 

To provide a sense of touch in the AR space, vibrotactile feedback was provided to the subjects through the Myo\texttrademark~armband (Thalmic Labs, Canada) when any contact of the vMPL was made with the other virtual objects in the scene. To control the vMPL, inertial sensors (see Section II-B) were placed on the upper arm and forearm to capture kinematic information. A total of 6 degrees of freedom (DOF) (3 for shoulder, 1 for elbow flexion/extension, 1 for wrist supination/pronation and 1 for hand open/close) of the vMPL could be controlled. While shoulder and elbow DOF were controlled using the inertial tracking, the wrist DOF was driven using pattern recognition on the subject's surface electromyogram (EMG) signals. EMG data was recorded and streamed through the Myo\texttrademark armband and pattern movements were decoded using linear discriminant analysis (LDA). As loading has shown to have an effect on pattern recognition systems, a dummy weight, with a mass equal to the physical Modular Prosthetic Limb, was wrapped around the upper limb while training on the HoloPHAM~\cite{betthauser2018limb},~\cite{cipriani2011influence}.

\begin{figure}[b]
      \centering
     \includegraphics[width=0.44\textwidth]{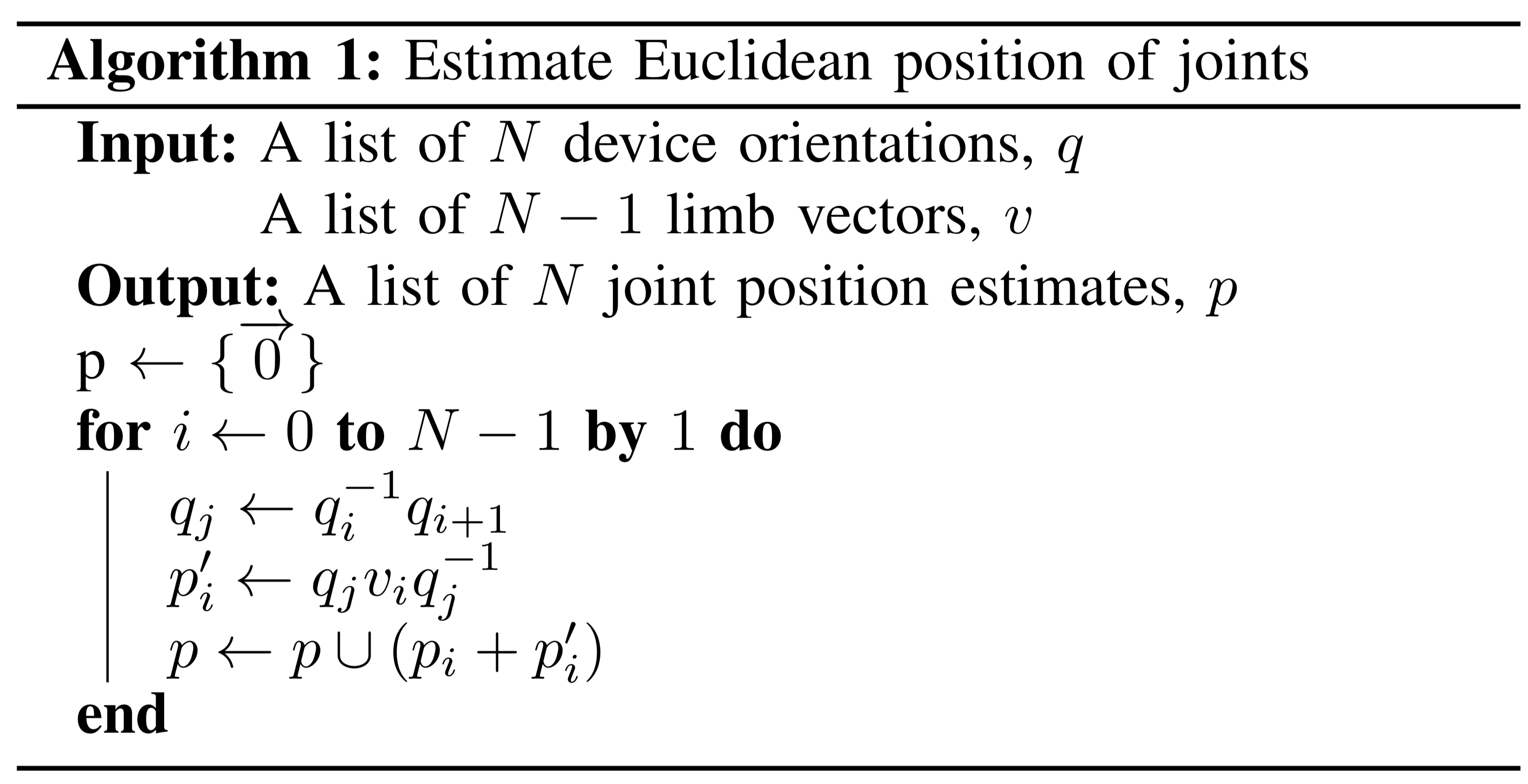}

   \end{figure} 
\subsection{Inertial Sensing \& Kinematic Assessment}
In order to both control the vMPL and track joint trajectories, a network of 9-axis MPU9250 (InvenSense, San Jose, CA) inertial measurement units~(IMUs) were used (see Fig. \ref{fig: HoloPHAM}). The three IMUs (on forearm, upperarm and chest) streamed absolute orientation values as quaternions, at 20~Hz, to a Bluetooth Low Energy (BLE) adapter, connected to a computer which computed Euler joint angles of the shoulder and the elbow. A Madgwick filter was used to efficiently filter and fuse raw sensor data based on the algorithm described here~\cite{madgwick2011estimation}. 

Besides controlling the vMPL, inertial sensing was used for kinematic comparison. Comparisons were made based on the rotational joint angle dynamics of the shoulder and elbow during a task. A baseline kinematic profile for each subject was used for comparison. The joint dynamics while performing the tasks on the PHAM at the beginning of the study was used as the baseline kinematic profile. For rotational joint angle dynamics, time series quaternions were compared using the dynamic time warping (DTW) algorithm. DTW is a popular mechanism for calculating similarity between two signals that may be independent of certain non-linear variations in the time dimension. DTW aligns the signals by warping the time dimension iteratively until an optimal warped path ($W$) is found. The detailed algorithm has been described here~\cite{berndt1994using}. DTW has been shown to be a valid method for quantifying movement quality in upper limb prosthesis use~\cite{thies2017skill}. For two quaternion time series $Q$ and $\hat{Q}$, of lengths $|Q|$ and $|\hat{Q}|$ respectively, an optimal warp path $W$ is constructed, with the distance of the warp path $W$ estimated by:
 \begin{equation}
     Dist(W) \quad = \quad \sum_{k=1}^{k=K}Dist(w_{ki},w_{kj})
 \end{equation}
 where $Dist(W)$ is the Euclidean distance of the warp path $W$, and $Dist(w_{ki},w_{kj})$ is the distance between the two quaternion point indexes (one from $Q$ and one from $\hat{Q}$) in the $k^{th}$ element of the warp path. A lower distance $Dist(W)$ value represents a greater similarity between two joint dynamics. Since the standard DTW algorithm runs in $O(N^2)$, we adopted a slightly optimized version known as FastDTW, which provides optimal alignments in $O(N)$ time and memory complexity~\cite{salvador2007toward}. 

To account for the metabolic cost involved during reaching tasks, energy expenditure, as an estimate of metabolic effort during reaching was computed~\cite{shadmehr2016representation}. The formulation estimates the energy expended to reach a distance $d$ and is given by:
\begin{equation}
    e_r = amT + b\frac{md^i}{T^{j-1}}
\end{equation}
where metabolic energy expenditure ($e_r$) is given as a function of the reach duration, $T$, reach distance, $d$, effective limb mass, $m$ and the parametric constants $a,b,i,j$. For reaching motor tasks, $a$ = 15, $b$ = 77, $i$ = 1.1, $j$ = 3. These values are adopted from previous research and have been estimated based on simulations of reaching movements at different distances and duration~\cite{shadmehr2016representation}. Reach distance \textit{d} is the Euclidean distance traversed by a joint in time \textit{T}. Estimation of Euclidean position of joints has been described here~\cite{hunt2018predictive}. Algorithm 1 describes the three dimensional joint position estimation based on compounding iteration.
   \begin{table}[b!]
\centering
\caption{Feature parameters of tasks}
\label{table}
\begin{tabular}{|l|l|l|l|l|l|l|l|l|}
\hline
Task No. ($T$) & 1   & 2   & 3 & 4 \\ \hline
$\Delta d$   & 1  &1   & -1 & 0 \\ \hline
$\Delta \theta$    & 0 & 1 & 0 & 0 \\ \hline
\end{tabular}
\end{table}
\subsection{Experiment}
\begin{figure}[t!]
      \centering
      \framebox{\parbox{3.15in}{      \includegraphics[width=0.44\textwidth]{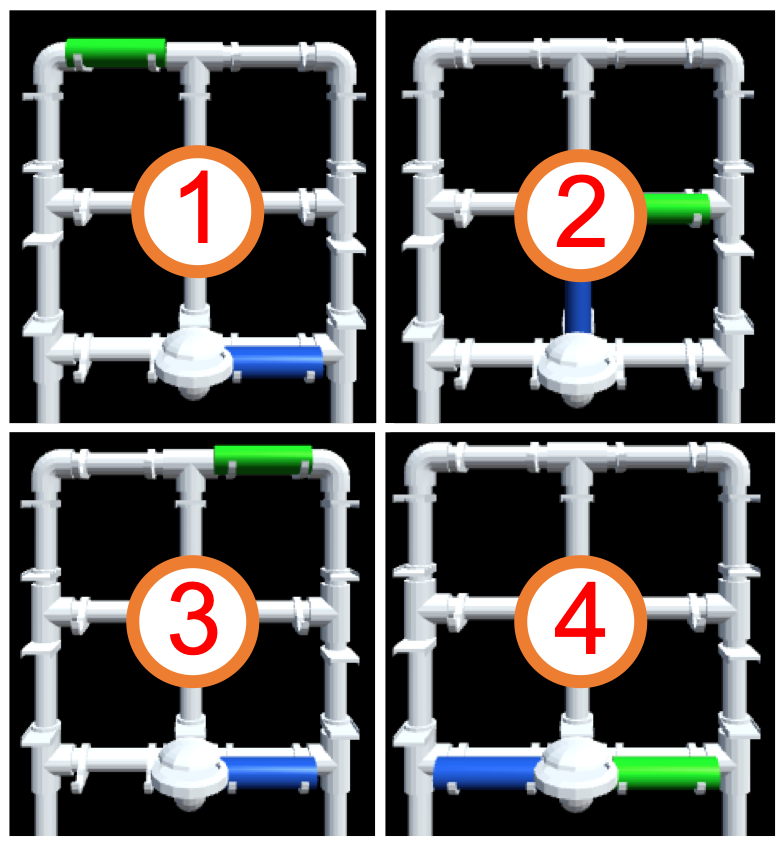}\\
      \includegraphics[width=0.44\textwidth]{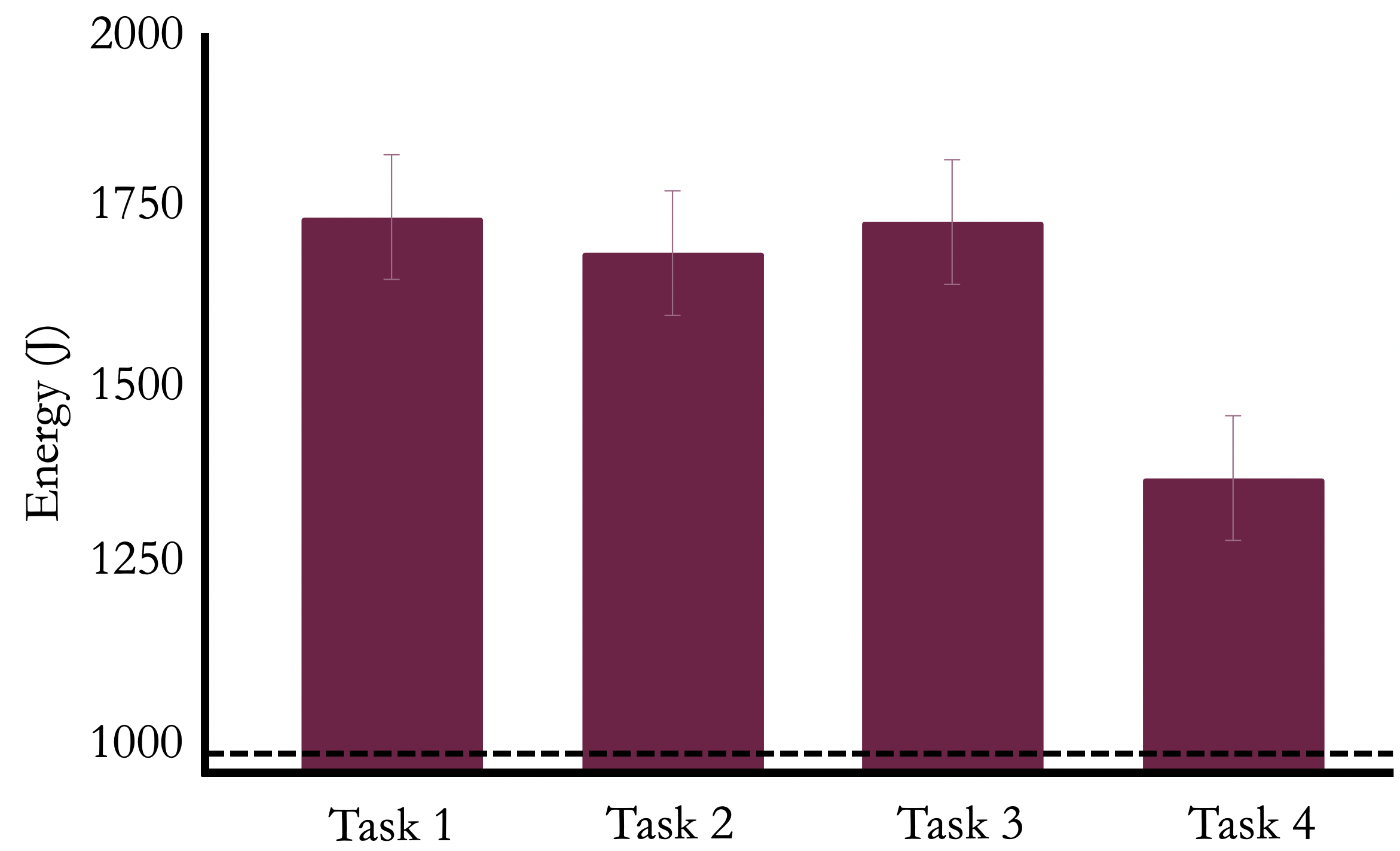}}}
      \caption{[Top] The four tasks on the PHAM and HoloPHAM involve reach and grasp movements which account for three dimensional kinematics and different grasp patterns. To ensure consistency, subjects stood at an arms length from the frame. The subjects were required to start and return from a default position (limbs fully extended to the sides of the trunk) during each task. The two dimensional relocation of primitive objects share feature parameters (see Table I) similar to the CRT. The target (in blue) location represents location from where the object has to be picked and the goal (in green) location represents the position where it has to be released. The order of the tasks were randomized during the experiment to prevent cognitive bias. [Bottom] Average metabolic cost for each of these tasks were computed based on equation 2. A baseline metabolic cost ($\approx$ 965 J) for reach and return has been represented by the black dashed line. The error bars represent the stand error of mean.}
      \label{fig: tasks}
   \end{figure} 
\subsubsection{Study Protocol}
The study was divided into three phases: 1) initial evaluation phase, 2) training phase, 3) testing phase and 4) washout phase. In the initial evaluation phase, the subjects were asked to perform a set of outcome measures that included the CRT, PHAM and HoloPHAM. For the purposes of analysis, task completion time, EMG and kinematic information were recorded. The training phase involved the subjects visiting a local clinic multiple times over 10 days, with no more than a 2-day gap between the sessions. In each visit, the subjects trained on the HoloPHAM, through a set of three repetitions of four tasks that were randomized during training.
After the training phase was over, in the testing phase, the subjects were asked to repeat the set of outcomes that were used in the initial evaluation phase. Post-training, a washout period of 5 days was involved were the subjects did not receive any training.

\subsubsection{Reach Tasks}
For depth perception in AR a series of reaching tasks with the vMPL was conducted, where the subjects (\textit{N=3}) had to guide the vMPL towards a goal endpoint position from a starting point in the AR space (see Fig. \ref{fig:Depth}a). In Fig. \ref{fig:Depth}a, $d_{\text{AR}}$ (in meters) denotes this distance value, where the starting point refers to the position where the vMPL is closest to the subject's chest and the endpoint refers to the position of a virtual cylinder in AR space. The cylinder was placed at the subject's chest level. Inertial trackers were used to estimate the endpoint of the physical hand (using Algorithm 1) and its reach distance was computed. For five set of trials, mean and standard deviation of the physical reach distance ($\mu_{phy}$ and $\sigma_{phy}$) was calculated. This experiment was conducted with three $d_{\text{AR}}$ values of 0.2 m, 0.25 m and 0.3 m, respectively. To ensure that only visual feedback is used to perceive depth, no additional vibrotactile feedback for touch was provided during this experiment. 
\subsubsection{PHAM \& HoloPHAM Tasks}
All tasks involved reach and grasp movements in a three dimensional space. While both the PHAM and HoloPHAM are capable of a diverse range of object manipulations, subjects were asked to complete a subset of tasks that resembled those of the CRT. Since tasks on these outcomes account for 2-dimensional movement and wrist rotation, a task $T(\Delta d, \Delta \theta)$ was parameterized by two features - 1) relative 2-D position $(\Delta d)$ and 2) relative angular position $(\Delta \theta)$  of the target location with respect to the initial location. Relative 2-D position determines whether the object has to be moved vertically $(\Delta d = -1)$, diagonally $(\Delta d = 1)$ or along the same level $(\Delta d = 0)$ from the initial position while the relative angular position describes the orientation of the target location with respect to the initial location and whether the subject would be required to rotate the end effector of the vMPL $(\Delta \theta = 1)$ or not $(\Delta \theta = 0)$. Table~\ref{table} shows the feature parametrization of all the tasks on the PHAM and HoloPHAM while Fig.\ref{fig: tasks} shows a visualization of the required manipulations. For all three outcome measures, there were three repetitions for each task. The sequence of tasks were randomized and subjects were allowed to complete them at their own pace in order to prevent cognitive bias and muscle fatigue, respectively.

\section{Results}
\subsection{Inertial Sensing Validation}

To enable efficient wireless kinematic tracking, IMUs and rigid body constraints were used to accurately estimate joint positions, up to the centimeter range. Wireless tracking through inertial sensors are known to be low cost, low weight and high accuracy solutions over optical and depth trackers. IMUs have become a popular kinematic tracker for human machine interfaces, especially for devices that require arm tracking capabilities~\cite{masters2015real}.
\subsubsection{Rotation}
\begin{figure}[h!]
      \centering
      \includegraphics[width=0.5\textwidth]{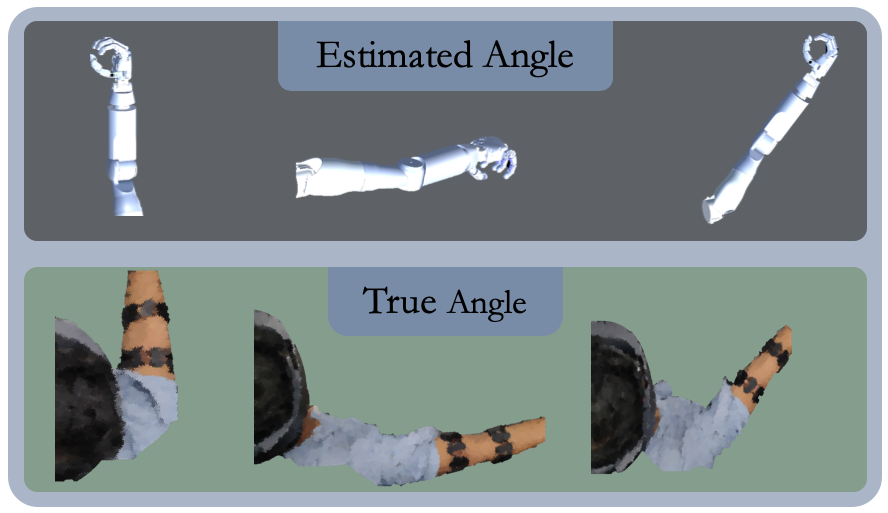}
      \caption{ Top view showing the estimated and true angles of the vMPL and actual physical limb rotation. Sensory accuracies were 3.38\degree~(1.34\degree precision) and 2.24\degree~(0.77\degree precision) for polar and azimuthal configurations, respectively. Inertial sensors were validated with a stepper motor as described here~\cite{hunt2018predictive}. A visual latency of the actual movement of the vMPL was observed to be between 500-800 ms. }
      \label{fig:SensorValid}
   \end{figure}

In our previous work we studied the angular (polar and azimuthal) accuracy of our kinematic tracking system~\cite{hunt2018predictive}. In our study we achieved a mean estimate accuracy of 2.81\degree~with 1.06\degree~precision, with a maximum endpoint hand tracking error of 7.06~cm (see Fig.\ref{fig:SensorValid}) due to angular perturbation. 
   \begin{figure}[b!]
      \centering
      \includegraphics[width=0.5\textwidth]{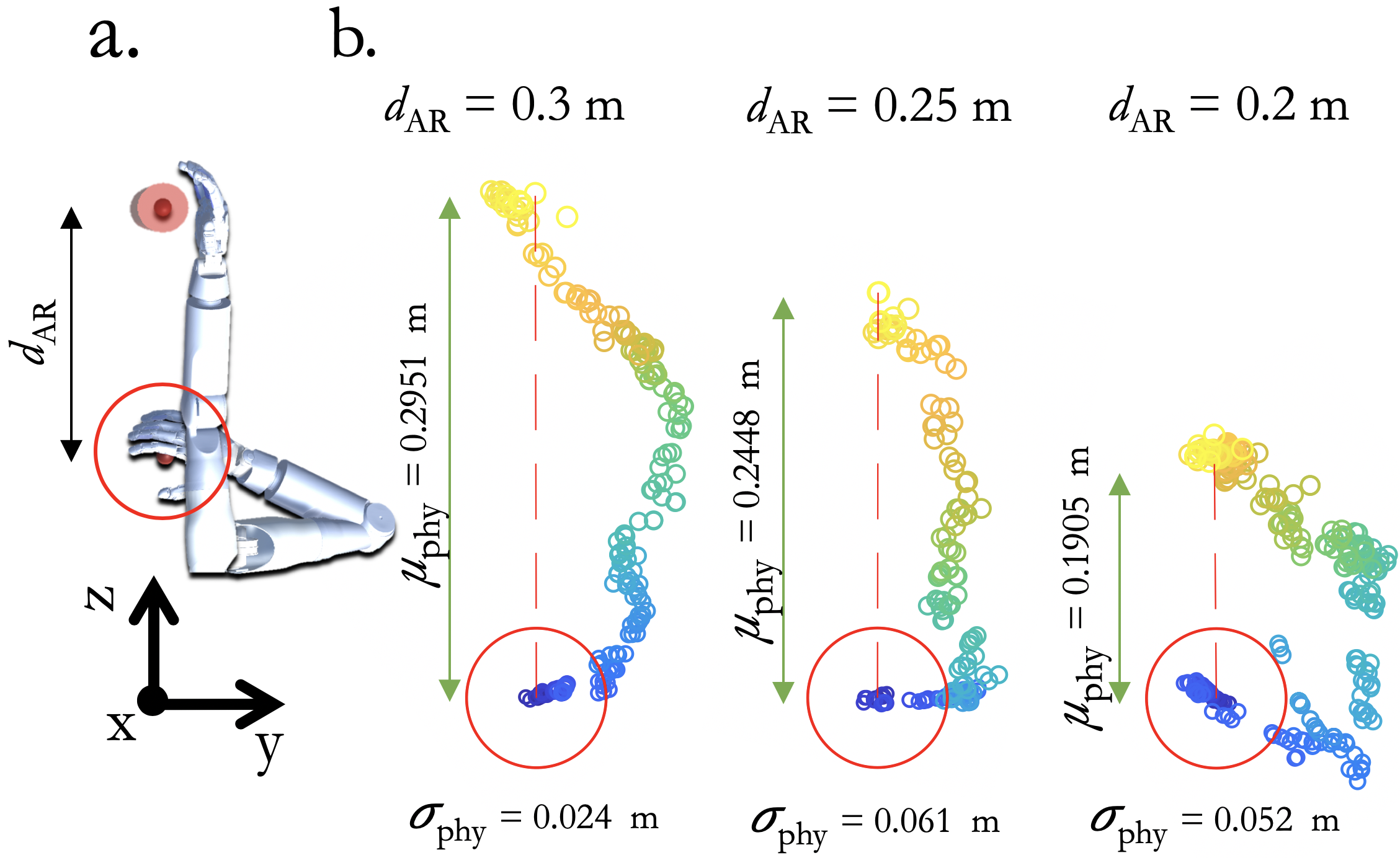}
      \caption{a. The configuration of the vMPL in the starting and end position during the target reach movement. Subjects were instructed to start from the same starting position (represented by the red circle) and reach towards the end point ( represented by the virtual red cylinder) with no time constraints. No vibrotactile feedback was provided during this experiment.$d_{\text{AR}}$ denotes the distance or depth of the cylinder from the starting position in the AR scene. b. Trajectory of the physical hand end-point, along the zy plane, is shown for multiple virtual depth values ($d_{\text{AR}}$). The color spectrum changes from blue to yellow from the starting till the end point. $\mu_{phy}$ denotes the mean distance traversed by the physical hand end-point for the corresponding $d_{\text{AR}}$, while $\sigma_{phy}$ denotes the standard deviation.  }
      \label{fig:Depth}
   \end{figure} 
\subsubsection{Translation}
To further validate inertial tracking, depth perception experiment was conducted (see Section II-C-2) where the virtual depth was compared with the physical hand displacement. In Fig. \ref{fig:Depth} we show the configuration of the experiment for the reach task and the trajectory of the physical endpoint along the zy plane, where the depth distance is along the x direction. With five set of trials with three subjects, a mean error in depth perception was 5.27 cm.

\subsubsection{Kinematics}
\begin{figure}[t!]
      \centering
          \includegraphics[width=0.5\textwidth]{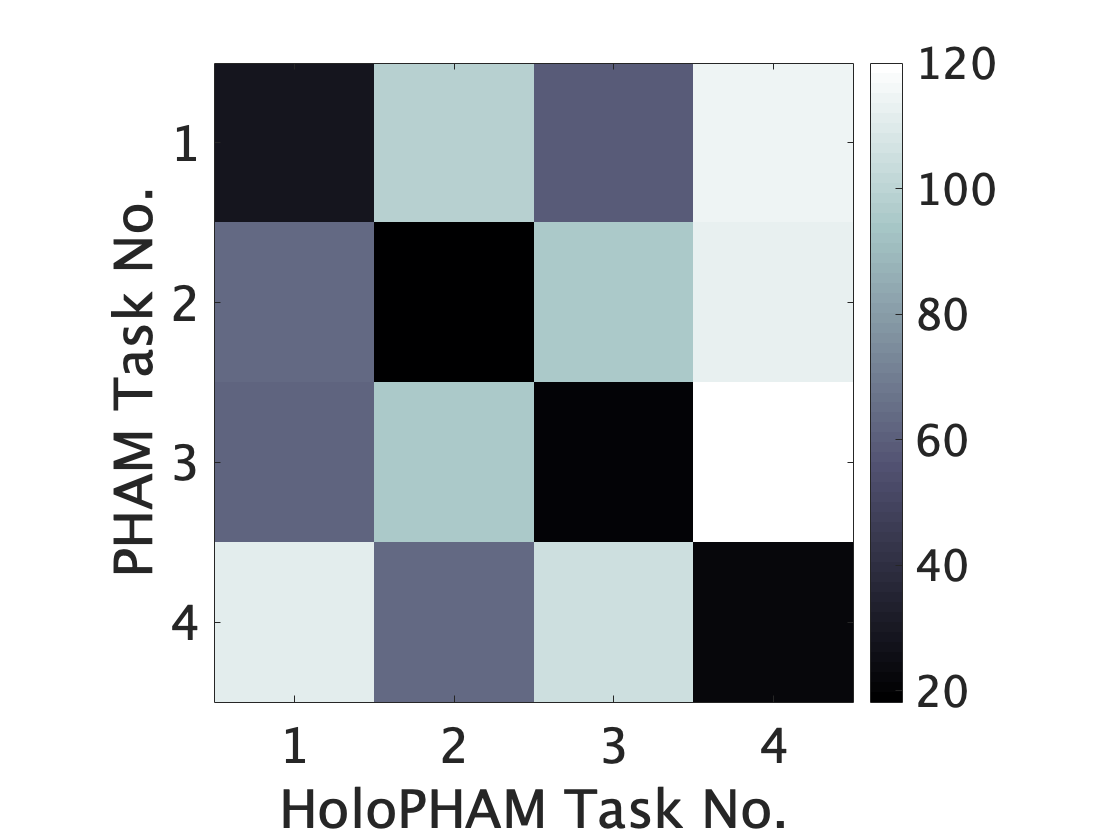}
       \caption{Confusion matrix showing the distances of the warped path ($C_{ij}$) between the tasks on our physical (PHAM, along y axis, $|Q_{i}|$) and augmented reality (HoloPHAM, along x axis, $|\hat{Q_j}|$) setup based on the DTW algorithm. Cost ($C_{ij}$) represents the mean (over five trials) measured using equation 1. A lower cost represents greater similarity between the joint dynamics in the two tasks.}
      \label{fig:confusion}
   \end{figure}

Ensuring that HoloPHAM accurately resembles reaching task of the PHAM, kinematic joint dynamics were compared in the two modalities. Since the HoloLens\texttrademark~uses built-in optical and inertial sensors for position and orientation tracking, calibration errors can result in mismatch between the physical and virtual space dynamics~\cite{qian2017comprehensive}. A similarity cost based on the DTW algorithm (see Eqn. 1) was used to compare the joint angle dynamics of the tasks performed on the PHAM and HoloPHAM (see Fig.\ref{fig:confusion}). 
   \begin{figure}[b!]
      \centering
          \includegraphics[width=0.5\textwidth]{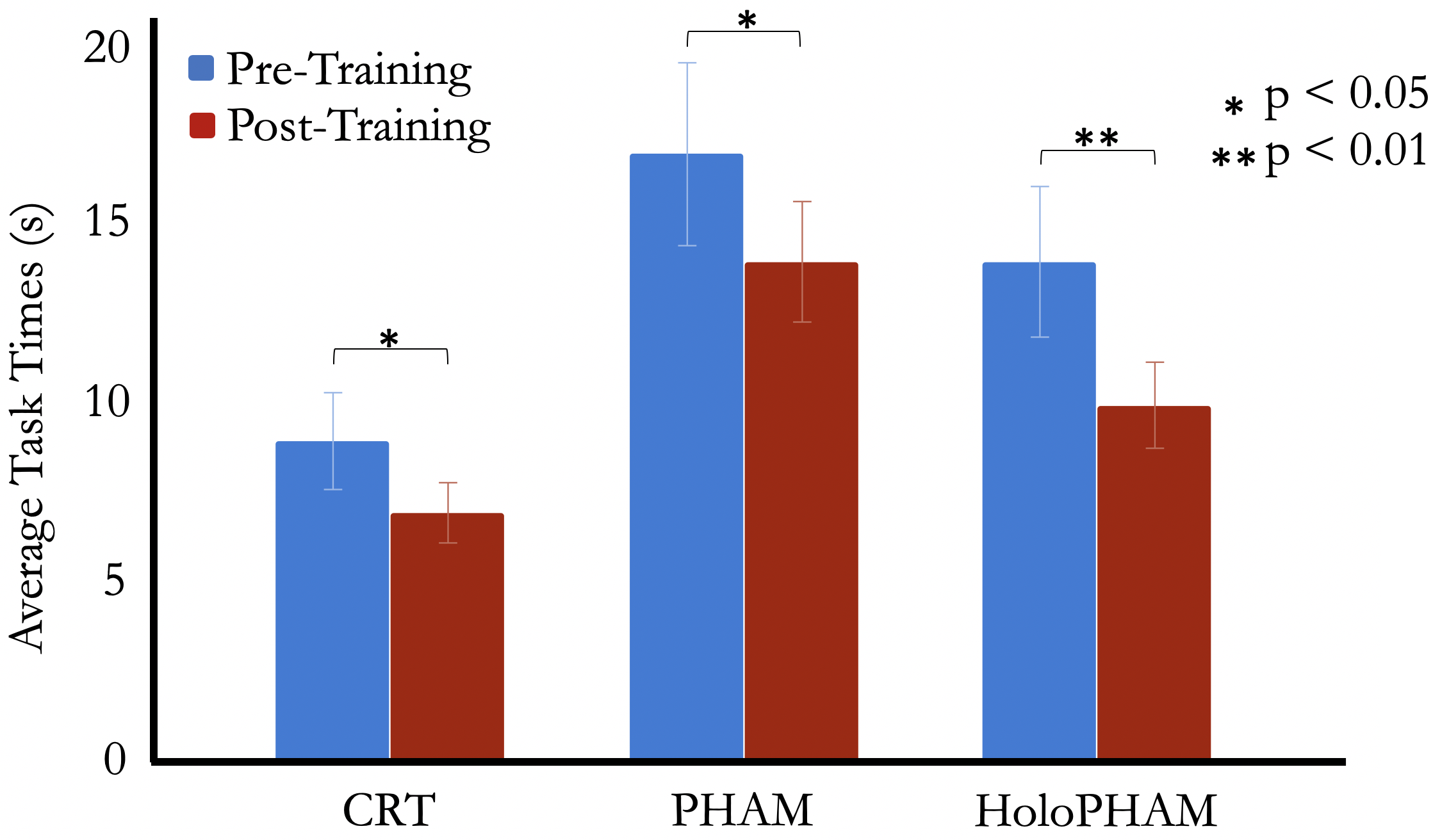}
      \caption{Task completion times for subjects (N=3, able-bodied) in the pre- and post-training phases are reported. * denotes the statistical significance at p \textless 0.05 and ** at p \textless 0.01.}
      \label{fig:taskTimes}
   \end{figure} 
\subsection{Effect of HoloPHAM Training}
After a training phase of 10 days, we observed the effects of performance during tasks on the CRT, PHAM and HoloPHAM, where we looked at the average completion times (see Fig.\ref{fig:taskTimes}). Statistically significant improvements were observed in both physical and virtual outcomes, with percentage reduction in completion times of 22.22\%, 23.5\% and 28.5\% in CRT, PHAM and HoloPHAM respectively.

   \begin{figure}[t!]
      \centering
          \includegraphics[width=0.5\textwidth]{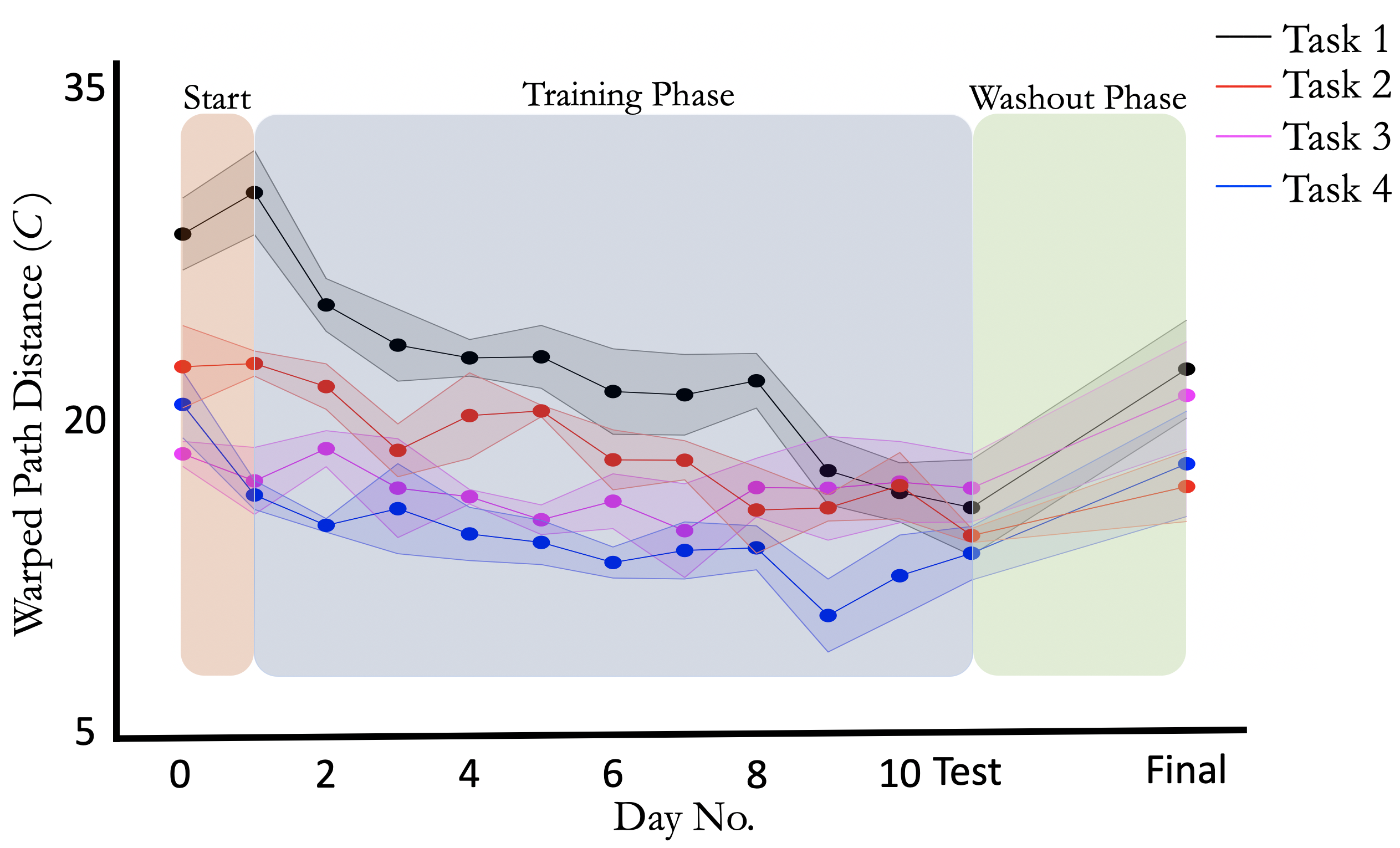}
      \caption{Progression of the warped path distance (similarity cost, $C$) of joint dynamics over the training phase across the four tasks on the HoloPHAM. The references (baselines) are the joint dynamics for each tasks measured on Day 0 with the tasks on the PHAM. Similarity cost after a washout period of 5 days, post test day, has also been reported. A lower warped path distance signifies greater similarity between the PHAM and HoloPHAM joint dynamics, for the same tasks ($C$ = 0 for two exactly same time series signals). The solid lines represent the average across all subjects (N=3), with the variance in data represented as shaded patches.  }
      \label{fig:costJointDynamics}
   \end{figure} 
During the training phase with the HoloPHAM, the joint dynamics during each task was recorded. In Fig. \ref{fig:costJointDynamics}, we show the mean and variance in similarity cost of the joint dynamics of HoloPHAM tasks with the corresponding baseline tasks on the PHAM. Comparison score was measured as the cost of the warped path using the DTW algorithm (see Eqn. 1). Cost from all sessions between the pre- and post-training phases are reported. There was a general reduction in the cost between pre- and post-training phase. After the washout phase, there was however, an increase in the cost across all tasks. 

\section{Discussion}
Upper limb prosthesis training involves repetition of reach and grasp movements that are designed to assist motor skill learning~\cite{simon2012patient}. Current state-of-the-art VR and AR prosthesis systems are limited as they focus only on a specific aspect of prosthesis control. In this work, we developed a multi-dimensional human-machine interaction system which offers several DOF control (both kinematic and myoelectric) of a virtual prosthesis. Prior to using our AR system for training, we conducted a series of validation experiments. These experiments were performed to understand the transfer of physical information into the AR space through our interface. To begin with, rotational and translational error in the interface were looked at (see Fig.\ref{fig:SensorValid} and \ref{fig:Depth}). To study how these variables affect the 3-dimensional reaching and grasp kinematics, joint dynamics were compared (see Fig.\ref{fig:confusion})
Corresponding reaching tasks on the PHAM and HoloPHAM showed greater similarity, suggesting that there is a linear correlation between the physical and AR space, in terms of kinematics. We then conducted a long-term study to investigate the effects of training with our AR system on standardized functional outcome measures. While completion times improved in all three measures between pre- and post-training (see Fig. \ref{fig:taskTimes}), the quality of motion in the HoloPHAM, in terms of joint dynamics, was also observed to converge towards the PHAM (see Fig.\ref{fig:costJointDynamics}). Improvements in task completion times suggests that training with our AR setup, by in itself, can enhance motor skill performance, with the effects, also translating to the physical outcome measures. Subsequent reduction in joint kinematic cost, while training with the HoloPHAM, suggests that subjects are able to develop a more intuitive control of the vMPL with time (see Fig.\ref{fig:costJointDynamics}). Our results show both quantitative and qualitative improvements in skill performance with  AR training. 

While motor learning processes are an internal phenomenon and cannot be directly observed, performance of subjects during skill experiments can be used an indicator for inferring these effects of training~\cite{edwards2010motor}. Such an internal model, believed to be based primarily in the cerebellar region of the brain, has been known to have synaptic plasticity, which can be trained and re-trained in a supervised manner~\cite{kawato1999internal}. Motor learning requires immense amount of sensation and perception~\cite{frank2016perceptual}. To facilitate learning, our AR system incorporated several aspects of these paradigms. The HoloPHAM enabled subjects to see virtual objects that are blended into the physical space, while also allowing them to maintain visual contact of the vMPL (see Fig. \ref{fig: HoloPHAM}). Visual contact of the hand and target objects has been known to provide assistive feedback for motor planning and control, allowing proper execution of the movement~\cite{churchill2000vision}.  Vibrotactile feedback allowed subjects to know when they made contact of virtual objects with the vMPL (see Fig. \ref{fig:touch}) . Previous research has shown that a non-invasive mechanical vibration on the upper limb stump, when coupled with a visual feedback of the virtual limb, stimulates the sensory-motor centers in the cortex~\cite{kurzynski2017computer}. Interpretation of sensory information and feedback enables the perception of the objects and visual cues in an environment. Accurate perception of virtual objects, within a range of few centimeters, was accomplished in our AR setup (see Fig. \ref{fig:Depth}). Lack of depth perception has been shown to negatively impact reaching performance, quality of movement as well grasping actions~\cite{bozzacchi2015lack}. Joint kinematic comparison of a physical limb with the virtual limb (vMPL in our case) throws light on motor coordination and the aspect of embodiment of virtual objects in AR (see Fig. \ref{fig:costJointDynamics}). The increase in joint kinematic cost, after the washout phase (see Fig.\ref{fig:costJointDynamics}), suggests that our system facilitated motor learning and illustrated the effect of practise for motor retention.

In our current system, a visual latency of the 3-dimensional movement of virtual limb was in the range of 500-800 ms. While larger than the real-time limit of 250 ms, our results suggest this system latency is still within the limits to facilitate motor learning. While a longer time study would be useful to see how the similarity between physical and virtual joint dynamics evolved, it essential to note that an exact similarity can be hard to achieve with the current technological limitations of current AR systems~\cite{radkowski2018augmented}. Future research in this space is directed towards observing the effects AR training has on amputee population's pattern recognition and kinematic performance. 
\section{Conclusion}
In this work we presented an AR based human-machine interaction system for prosthesis control. The system provided both visual and vibrotactile feedback to the subjects during interaction. Our AR setup, based upon an exisiting physical functional ouctome measure, allowed both myoelectric and kinematic control of a virtual prosthesis, with several degrees of freedom. Kinematic relationship between the physical and AR space were studied. Furthermore, the effects of training with our AR setup over a period of 10 days were investigated. Both quantitative and qualitative aspects of reach and object manipulation performance showed improvements. These effects depreciated, but were still retained, after a washout phase without training, suggesting that our AR setup facilitated motor skill learning. Future work will investigate the effects on a broader subject population (both amputee and able-bodied) over a longer study duration. 
\section{Acknowledgment}
We wish to acknowledge Johns Hopkins Applied Physics Laboratory (JHU/APL) for developing and making available the Virtual Integration Environment (VIE), which the virtual Modular Prosthesis Limb (vMPL) is a part of. The VIE was developed under the Revolutionizing Prosthetics program. The authors would like to thank the human subjects who participated in this study and Johns Hopkins University Applied Physics Laboratory for providing the Microsoft HoloLens\texttrademark. 
\bibliography{IEEEabrv.bib,Bibliography.bib}{}

\begin{thebibliography}{10}
\providecommand{\url}[1]{#1}
\csname url@samestyle\endcsname
\providecommand{\newblock}{\relax}
\providecommand{\bibinfo}[2]{#2}
\providecommand{\BIBentrySTDinterwordspacing}{\spaceskip=0pt\relax}
\providecommand{\BIBentryALTinterwordstretchfactor}{4}
\providecommand{\BIBentryALTinterwordspacing}{\spaceskip=\fontdimen2\font plus
\BIBentryALTinterwordstretchfactor\fontdimen3\font minus
  \fontdimen4\font\relax}
\providecommand{\BIBforeignlanguage}[2]{{%
\expandafter\ifx\csname l@#1\endcsname\relax
\typeout{** WARNING: IEEEtran.bst: No hyphenation pattern has been}%
\typeout{** loaded for the language `#1'. Using the pattern for}%
\typeout{** the default language instead.}%
\else
\language=\csname l@#1\endcsname
\fi
#2}}
\providecommand{\BIBdecl}{\relax}
\BIBdecl

\bibitem{kumahara2004daily}
H.~Kumahara, H.~Tanaka, and Y.~Schutz, ``Daily physical activity assessment:
  what is the importance of upper limb movements vs whole body movements?''
  \emph{International journal of obesity}, vol.~28, no.~9, p. 1105, 2004.

\bibitem{esquenazi2004amputation}
A.~Esquenazi, ``Amputation rehabilitation and prosthetic restoration. from
  surgery to community reintegration,'' \emph{Disability and rehabilitation},
  vol.~26, no. 14-15, pp. 831--836, 2004.

\bibitem{johnson2014prosthetic}
S.~S. Johnson and E.~Mansfield, ``Prosthetic training: upper limb,''
  \emph{Physical Medicine and Rehabilitation Clinics}, vol.~25, no.~1, pp.
  133--151, 2014.

\bibitem{malone1984immediate}
J.~Malone, L.~Fleming, J.~Roberson, J.~T. Whitesides, J.~Leal, J.~Poole, and
  R.~Grodin, ``Immediate, early, and late postsurgical management of upper-limb
  amputation.'' \emph{Journal of rehabilitation research and development},
  vol.~21, no.~1, pp. 33--41, 1984.

\bibitem{biddiss2007upper}
E.~A. Biddiss and T.~T. Chau, ``Upper limb prosthesis use and abandonment: a
  survey of the last 25 years,'' \emph{Prosthet. Orthot. Int.}, vol.~31, no.~3,
  pp. 236--257, 2007.

\bibitem{wheaton2017neurorehabilitation}
L.~A. Wheaton, ``Neurorehabilitation in upper limb amputation: understanding
  how neurophysiological changes can affect functional rehabilitation,''
  \emph{Journal of neuroengineering and rehabilitation}, vol.~14, no.~1, p.~41,
  2017.

\bibitem{wright2009prosthetic}
V.~Wright, ``Prosthetic outcome measures for use with upper limb amputees: A
  systematic review of the peer-reviewed literature, 1970 to 2009,'' \emph{JPO:
  Journal of Prosthetics and Orthotics}, vol.~21, no.~9, pp. P3--P63, 2009.

\bibitem{prince2008comparison}
S.~A. Prince, K.~B. Adamo, M.~E. Hamel, J.~Hardt, S.~C. Gorber, and
  M.~Tremblay, ``A comparison of direct versus self-report measures for
  assessing physical activity in adults: a systematic review,''
  \emph{International Journal of Behavioral Nutrition and Physical Activity},
  vol.~5, no.~1, p.~56, 2008.

\bibitem{hill2009functional}
W.~Hill, {\O}.~Stavdahl, L.~N. Hermansson, P.~Kyberd, S.~Swanson, and
  S.~Hubbard, ``Functional outcomes in the who-icf model: establishment of the
  upper limb prosthetic outcome measures group,'' \emph{JPO: Journal of
  Prosthetics and Orthotics}, vol.~21, no.~2, pp. 115--119, 2009.

\bibitem{bouwsema2012determining}
H.~Bouwsema, P.~J. Kyberd, W.~Hill, C.~K. van~der Sluis, and R.~M. Bongers,
  ``Determining skill level in myoelectric prosthesis use with multiple outcome
  measures,'' \emph{J Rehabil Res Dev}, vol.~49, no.~9, pp. 1331--1348, 2012.

\bibitem{bongers2012bernstein}
R.~M. Bongers, P.~J. Kyberd, H.~Bouwsema, L.~P. Kenney, D.~H. Plettenburg, and
  C.~K. Van~der Sluis, ``Bernstein’s levels of construction of movements
  applied to upper limb prosthetics,'' \emph{JPO: Journal of Prosthetics and
  Orthotics}, vol.~24, no.~2, pp. 67--76, 2012.

\bibitem{major2014comparison}
M.~J. Major, R.~L. Stine, C.~W. Heckathorne, S.~Fatone, and S.~A. Gard,
  ``Comparison of range-of-motion and variability in upper body movements
  between transradial prosthesis users and able-bodied controls when executing
  goal-oriented tasks,'' \emph{Journal of neuroengineering and rehabilitation},
  vol.~11, no.~1, p. 132, 2014.

\bibitem{thies2017skill}
S.~B. Thies, L.~P. Kenney, M.~Sobuh, A.~Galpin, P.~Kyberd, R.~Stine, and M.~J.
  Major, ``Skill assessment in upper limb myoelectric prosthesis users:
  Validation of a clinically feasible method for characterising upper limb
  temporal and amplitude variability during the performance of functional
  tasks,'' \emph{Medical engineering \& physics}, vol.~47, pp. 137--143, 2017.

\bibitem{metzger2012characterization}
A.~J. Metzger, A.~W. Dromerick, R.~J. Holley, and P.~S. Lum, ``Characterization
  of compensatory trunk movements during prosthetic upper limb reaching
  tasks,'' \emph{Archives of physical medicine and rehabilitation}, vol.~93,
  no.~11, pp. 2029--2034, 2012.

\bibitem{hunt2011pham}
C.~Hunt, R.~Yerrabelli, C.~Clancy, L.~Osborn, R.~Kaliki, and N.~Thakor, ``Pham:
  prosthetic hand assessment measure,'' in \emph{Proc. Myoelec. Controls
  Symp.}, 2017, p. 221.

\bibitem{dromerick2008effect}
A.~W. Dromerick, C.~N. Schabowsky, R.~J. Holley, B.~Monroe, A.~Markotic, and
  P.~S. Lum, ``Effect of training on upper-extremity prosthetic performance and
  motor learning: a single-case study,'' \emph{Archives of physical medicine
  and rehabilitation}, vol.~89, no.~6, pp. 1199--1204, 2008.

\bibitem{simon2012patient}
A.~M. Simon, B.~A. Lock, and K.~A. Stubblefield, ``Patient training for
  functional use of pattern recognition--controlled prostheses,'' \emph{Journal
  of prosthetics and orthotics: JPO}, vol.~24, no.~2, p.~56, 2012.

\bibitem{hargrove2007real}
L.~Hargrove, Y.~Losier, B.~Lock, K.~Englehart, and B.~Hudgins, ``A real-time
  pattern recognition based myoelectric control usability study implemented in
  a virtual environment,'' in \emph{Engineering in Medicine and Biology
  Society, 2007. EMBS 2007. 29th Annual International Conference of the
  IEEE}.\hskip 1em plus 0.5em minus 0.4em\relax IEEE, 2007, pp. 4842--4845.

\bibitem{simon2011target}
A.~M. Simon, L.~J. Hargrove, B.~A. Lock, and T.~A. Kuiken, ``The target
  achievement control test: Evaluating real-time myoelectric pattern
  recognition control of a multifunctional upper-limb prosthesis,''
  \emph{Journal of rehabilitation research and development}, vol.~48, no.~6, p.
  619, 2011.

\bibitem{lambrecht2011virtual}
J.~M. Lambrecht, C.~L. Pulliam, and R.~F. Kirsch, ``Virtual reality environment
  for simulating tasks with a myoelectric prosthesis: an assessment and
  training tool,'' \emph{Journal of prosthetics and orthotics: JPO}, vol.~23,
  no.~2, p.~89, 2011.

\bibitem{van2016task}
L.~van Dijk, C.~K. van~der Sluis, H.~W. van Dijk, and R.~M. Bongers,
  ``Task-oriented gaming for transfer to prosthesis use,'' \emph{IEEE
  Transactions on Neural Systems and Rehabilitation Engineering}, vol.~24,
  no.~12, pp. 1384--1394, 2016.

\bibitem{howard2017meta}
M.~C. Howard, ``A meta-analysis and systematic literature review of virtual
  reality rehabilitation programs,'' \emph{Computers in Human Behavior},
  vol.~70, pp. 317--327, 2017.

\bibitem{powell2014user}
M.~A. Powell, R.~R. Kaliki, and N.~V. Thakor, ``User training for pattern
  recognition-based myoelectric prostheses: Improving phantom limb movement
  consistency and distinguishability,'' \emph{IEEE Transactions on Neural
  Systems and Rehabilitation Engineering}, vol.~22, no.~3, pp. 522--532, 2014.

\bibitem{hargrove2018control}
L.~Hargrove, L.~Miller, K.~Turner, and T.~Kuiken, ``Control within a virtual
  environment is correlated to functional outcomes when using a physical
  prosthesis,'' \emph{Journal of neuroengineering and rehabilitation}, vol.~15,
  no.~1, p.~60, 2018.

\bibitem{brooks1999s}
F.~P. Brooks, ``What's real about virtual reality?'' \emph{IEEE Computer
  graphics and applications}, vol.~19, no.~6, pp. 16--27, 1999.

\bibitem{ebenholtz1992motion}
S.~M. Ebenholtz, ``Motion sickness and oculomotor systems in virtual
  environments,'' \emph{Presence: Teleoperators \& Virtual Environments},
  vol.~1, no.~3, pp. 302--305, 1992.

\bibitem{wann1997health}
J.~P. Wann and M.~Mon-Williams, ``Health issues with virtual reality displays:
  What we do know and what we don't,'' \emph{ACM SIGGRAPH Computer Graphics},
  vol.~31, no.~2, pp. 53--57, 1997.

\bibitem{burke2010augmented}
J.~W. Burke, M.~McNeill, D.~Charles, P.~J. Morrow, J.~Crosbie, and
  S.~McDonough, ``Augmented reality games for upper-limb stroke
  rehabilitation,'' in \emph{2010 Second International Conference on Games and
  Virtual Worlds for Serious Applications}.\hskip 1em plus 0.5em minus
  0.4em\relax IEEE, 2010, pp. 75--78.

\bibitem{luo2006integration}
X.~Luo, T.~Kline, H.~C. Fischer, K.~A. Stubblefield, R.~V. Kenyon, and D.~G.
  Kamper, ``Integration of augmented reality and assistive devices for
  post-stroke hand opening rehabilitation,'' in \emph{2005 IEEE Engineering in
  Medicine and Biology 27th Annual Conference}.\hskip 1em plus 0.5em minus
  0.4em\relax IEEE, 2006, pp. 6855--6858.

\bibitem{alamri2010ar}
A.~Alamri, J.~Cha, and A.~El~Saddik, ``Ar-rehab: An augmented reality framework
  for poststroke-patient rehabilitation,'' \emph{IEEE Transactions on
  Instrumentation and Measurement}, vol.~59, no.~10, pp. 2554--2563, 2010.

\bibitem{dunn2017virtual}
J.~Dunn, E.~Yeo, P.~Moghaddampour, B.~Chau, and S.~Humbert, ``Virtual and
  augmented reality in the treatment of phantom limb pain: a literature
  review,'' \emph{NeuroRehabilitation}, vol.~40, no.~4, pp. 595--601, 2017.

\bibitem{ortiz2014treatment}
M.~Ortiz-Catalan, N.~Sander, M.~B. Kristoffersen, B.~H{\aa}kansson, and
  R.~Br{\aa}nemark, ``Treatment of phantom limb pain (plp) based on augmented
  reality and gaming controlled by myoelectric pattern recognition: a case
  study of a chronic plp patient,'' \emph{Frontiers in neuroscience}, vol.~8,
  p.~24, 2014.

\bibitem{sharmaAR}
A.~Sharma, C.~Hunt, A.~Maheshwari, L.~Osborn, G.~Levay, R.~Kaliki, A.~Soares,
  and N.~Thakor, ``A mixed- reality training environment for upper limb
  prosthesis control,'' in \emph{2018 IEEE Biomedical Circuits asnd
  Systems}.\hskip 1em plus 0.5em minus 0.4em\relax IEEE, 2018, pp. 1--4.

\bibitem{kyberd2018characterisation}
P.~Kyberd, A.~Hussaini, and G.~Maillet, ``Characterisation of the clothespin
  relocation test as a functional assessment tool,'' \emph{Journal of
  Rehabilitation and Assistive Technologies Engineering}, vol.~5, p.
  2055668317750810, 2018.

\bibitem{betthauser2018limb}
J.~L. Betthauser, C.~L. Hunt, L.~E. Osborn, M.~R. Masters, G.~L{\'e}vay, R.~R.
  Kaliki, and N.~V. Thakor, ``Limb position tolerant pattern recognition for
  myoelectric prosthesis control with adaptive sparse representations from
  extreme learning,'' \emph{IEEE Transactions on Biomedical Engineering},
  vol.~65, no.~4, pp. 770--778, 2018.

\bibitem{cipriani2011influence}
C.~Cipriani, R.~Sassu, M.~Controzzi, and M.~C. Carrozza, ``Influence of the
  weight actions of the hand prosthesis on the performance of pattern
  recognition based myoelectric control: preliminary study,'' in
  \emph{Engineering in Medicine and Biology Society, EMBC, 2011 Annual
  International Conference of the IEEE}.\hskip 1em plus 0.5em minus 0.4em\relax
  IEEE, 2011, pp. 1620--1623.

\bibitem{madgwick2011estimation}
S.~O. Madgwick, A.~J. Harrison, and R.~Vaidyanathan, ``Estimation of imu and
  marg orientation using a gradient descent algorithm,'' in
  \emph{Rehabilitation Robotics (ICORR), 2011 IEEE International Conference
  on}.\hskip 1em plus 0.5em minus 0.4em\relax IEEE, 2011, pp. 1--7.

\bibitem{berndt1994using}
D.~J. Berndt and J.~Clifford, ``Using dynamic time warping to find patterns in
  time series.'' in \emph{KDD workshop}, vol.~10, no.~16.\hskip 1em plus 0.5em
  minus 0.4em\relax Seattle, WA, 1994, pp. 359--370.

\bibitem{salvador2007toward}
S.~Salvador and P.~Chan, ``Toward accurate dynamic time warping in linear time
  and space,'' \emph{Intelligent Data Analysis}, vol.~11, no.~5, pp. 561--580,
  2007.

\bibitem{shadmehr2016representation}
R.~Shadmehr, H.~J. Huang, and A.~A. Ahmed, ``A representation of effort in
  decision-making and motor control,'' \emph{Current biology}, vol.~26, no.~14,
  pp. 1929--1934, 2016.

\bibitem{hunt2018predictive}
C.~L. Hunt, A.~Sharma, L.~E. Osborn, R.~R. Kaliki, and N.~V. Thakor,
  ``Predictive trajectory estimation during rehabilitative tasks in augmented
  reality using inertial sensors,'' in \emph{2018 IEEE Biomedical Circuits and
  Systems Conference (BioCAS)}.\hskip 1em plus 0.5em minus 0.4em\relax IEEE,
  2018, pp. 1--4.

\bibitem{masters2015real}
M.~Masters, L.~Osborn, N.~Thakor, and A.~Soares, ``Real-time arm tracking for
  hmi applications,'' in \emph{Wearable and Implantable Body Sensor Networks
  (BSN), 2015 IEEE 12th International Conference on}.\hskip 1em plus 0.5em
  minus 0.4em\relax IEEE, 2015, pp. 1--4.

\bibitem{qian2017comprehensive}
L.~Qian, E.~Azimi, P.~Kazanzides, and N.~Navab, ``Comprehensive tracker based
  display calibration for holographic optical see-through head-mounted
  display,'' \emph{arXiv preprint arXiv:1703.05834}, 2017.

\bibitem{edwards2010motor}
W.~H. Edwards, \emph{Motor learning and control: From theory to
  practice}.\hskip 1em plus 0.5em minus 0.4em\relax Cengage Learning, 2010.

\bibitem{kawato1999internal}
M.~Kawato, ``Internal models for motor control and trajectory planning,''
  \emph{Current opinion in neurobiology}, vol.~9, no.~6, pp. 718--727, 1999.

\bibitem{frank2016perceptual}
C.~Frank, W.~M. Land, and T.~Schack, ``Perceptual-cognitive changes during
  motor learning: The influence of mental and physical practice on mental
  representation, gaze behavior, and performance of a complex action,''
  \emph{Frontiers in psychology}, vol.~6, p. 1981, 2016.

\bibitem{churchill2000vision}
A.~Churchill, B.~Hopkins, L.~R{\"o}nnqvist, and S.~Vogt, ``Vision of the hand
  and environmental context in human prehension,'' \emph{Experimental Brain
  Research}, vol. 134, no.~1, pp. 81--89, 2000.

\bibitem{kurzynski2017computer}
M.~Kurzynski, A.~Jaskolska, J.~Marusiak, A.~Wolczowski, P.~Bierut,
  L.~Szumowski, J.~Witkowski, and K.~Kisiel-Sajewicz, ``Computer-aided training
  sensorimotor cortex functions in humans before the upper limb transplantation
  using virtual reality and sensory feedback,'' \emph{Comput. Biol. Med.},
  vol.~87, pp. 311--321, 2017.

\bibitem{bozzacchi2015lack}
C.~Bozzacchi and F.~Domini, ``Lack of depth constancy for grasping movements in
  both virtual and real environments,'' \emph{Journal of neurophysiology}, vol.
  114, no.~4, p. 2242, 2015.

\bibitem{radkowski2018augmented}
R.~Radkowski and S.~Kanunganti, ``Augmented reality system calibration for
  assembly support with the microsoft hololens,'' in \emph{ASME 2018 13th
  International Manufacturing Science and Engineering Conference}.\hskip 1em
  plus 0.5em minus 0.4em\relax American Society of Mechanical Engineers, 2018,
  pp. V003T02A021--V003T02A021.

\end{thebibliography}
\bibliographystyle{IEEEtran}

\end{document}